\documentclass[prd,onecolumn,tightenlines]{revtex4}
\usepackage{amsmath}
\usepackage{amssymb}
\usepackage{graphicx}

\newcommand{\be}{\begin{eqnarray}}
\newcommand{\ee}{\end{eqnarray}}

 \newcommand{\gsim}{\mathrel{\hbox{\rlap{\lower.55ex \hbox {$\sim$}}
                   \kern-.3em \raise.4ex \hbox{$>$}}}}
\newcommand{\lsim}{\mathrel{\hbox{\rlap{\lower.55ex \hbox {$\sim$}}
                   \kern-.3em \raise.4ex \hbox{$<$}}}}

\begin{document}

\title{Large transverse momentum dilepton production in heavy ion collisions with two-photon processes }
\author{ Yong-Ping Fu$^{1)}$, and Yun-De Li$^{2)}$ \\ Department of Physics, Yunnan University, Kunming 650091, China\\
1) ynufyp@sina.cn; 2) yndxlyd@163.com }

\begin{abstract}
The cold component of large transverse momentum dilepton production
via semi-coherent two-photon interaction is calculated. The cold
contribution is essential to the dilepton spectra in the soft region
for different mass bins. The results are compared with the PHENIX
experimental data at RHIC, and we find that the modification of
semi-coherent two-photon processes is more evident with the rising
dilepton mass bins.


\end{abstract}

\maketitle

\address {Department of Physics, Yunnan University, Kunming 650091, People's
Republic of China }

\section{Introduction}

The main propose of ultrarelativistic heavy ion collisions is to
probe the thermal information of the dense thermalized matter which
is named quark-gluon plasma (QGP).The electromagnetic radiation
emitted from the center of the collision is a kind of very clean
information due to the real and virtual photons which do not
interact strongly, therefore photons and dileptons can escape from
the dense medium without strong interactions, and may carry the
information of the center dynamics\cite{1,2,3,4,5,6,7,8,9,10}.
However, so far no evident experiments show that some information is
exactly produced from the thermalized
medium\cite{11,12,13,14,15,16,17,18,19,20,21}.

The $\rho$ meson couples to the $\pi^{+}\pi^{-}$ channel strongly,
and the lifetime of $\rho$ meson is much shorter than the one of the
expected hot hadronic gas. The yield of the $\rho$ mess spectrum may
be modified in the hot medium due to the chiral symmetry
restoration. The scenario of mass dropping or melting in a hot
medium successfully interprets the dilepton yield enhancement in the
low mass region at Super Proton Synchrotron (SPS) for Pb-Au 158
GeV/A collisions\cite{22,23,24,25,26,27,28}. Recently, the
measurement of the dilepton continuum at Relativistic Heavy Ion
Collider (RHIC) energies has been performed by the PHENIX
experiments for Au-Au 200 GeV/A collisions\cite{29,30}. The dilepton
yield in the low mass range between 0.2 and 0.8 GeV is enhanced by a
factor of 2$\sim$3 compared with the expectation from hadron decays.
However, such a modifying scenario can not well explain the
enhancement in the Au-Au collisions at RHIC. The imperfect modifying
models of hadron decays present other probable mechanisms to explain
the enhancement of the dilepton yield at low mass. Moreover, the
dilepton enhancement at RHIC is implied that such phenomena are
related strongly to the hot plasma scenario.

The contribution of thermal dileptons at the low mass is covered by
the cocktail of hadron decays because the vector meson peaks is more
pronounced than the thermal spectrum. The thermal information is
dominant in the intermediate mass region between the $\phi$ and
$J/\Psi$ vector meson for the phase transition theory, but the
contribution of dileptons in this region also can be explained by
the decays of charmed mesons. The NA60 collaboration has also
observed an enhancement at intermediate mass. The data suggest that
such an enhancement may include a thermal information and not just
charm decays\cite{31}.

Except for the above problems, a new puzzle was performed by PHENIX
collaboration \cite{32}. The $P_{T}$ spectra of dileptons in $p+p$
collisions for different mass bins are in agreement with the
expectation of the cocktail, charm decays and direct contributions.
The agreement also exists in the Au-Au collisions at high $P_{T}$ (
$P_{T}>1$GeV). The excess contribution above the cocktail and charm
decays is consistent with the contribution from direct dileptons. In
the soft region of $P_{T}<1$GeV, pQCD is out of use, and the spectra
are still higher than the expectation of cocktail and charm decays.
If fitted with exponential in this soft window, one has a slope
$T_{eff}\approx 100$MeV, which is almost twice smaller than the
typical slope of the standard thermal component. Another candidate
in the soft window is the new "cold" component which is still valid
in the low $P_{T}$ region. In Ref \cite{33} the authors suggest a
new cold dilepton production mechanism which is not discussed in the
standard theory list before.

\section{General formalism}

The semi-coherent two-photon processes are the so-called dilepton
source for small mass bins and low $P_{T}$. Shuryak E. et. al. use
the equivalent photon approximation to determine the differential
cross section for the $\gamma\gamma$ processes in Au-Au collisions,
and have concluded that the semi-coherent production of dileptons
does not contribute significantly to the PHENIX data\cite{33}.
However,we derive the yield of dileptons for $\gamma\gamma$
processes with the restriction $b\ll 2R_{\perp}$, and finally find
that the production of double photon interaction has a positive
contribution to the cocktail and charm decays in the soft region.
One notes that the impact parameter $b$ is smaller than the
transverse dimension of the system $R_{\perp}$, and that the dense
hot medium can not be created in the heavy ion collisions if $b>
2R_{\perp}$ because the QGP is centrality dependent.

The equivalent photon spectrum corresponding to a point charge $Ze$,
moving with a velocity $v$ is given by
\begin{eqnarray}
n(\omega)=Z^{2}\frac{2\alpha}{\pi v^{2}}[\xi
K_{0}K_{1}-\frac{v^{2}\xi^{2}}{2}(K_{1}^{2}-K_{0}^{2})],
\end{eqnarray}
where the argument of the modified Bessel functions is $\xi=\omega
R_{min}/\gamma v$, and $R_{min}$ corresponds to the radius of the
radiation system with the maximum energy of the photons. After the
MacDonald approximation the above formulation can be written into a
compact form as\cite{34,35}
\begin{eqnarray}
n(\omega)=\frac{2Z^{2}\alpha}{\pi}\mathrm{ln} \frac{\gamma}{\omega
R_{min}},
\end{eqnarray}
where $R_{min}\sim 7 fm$ and $\gamma=106$ for Au-Au 200GeV/A
collisions at RHIC. In this article we use the natural units, namely
$h=c=1$. The differential cross section of two-photon interaction
for Au-Au collisions is given by
\begin{eqnarray}
d\sigma=\sigma_{\gamma \gamma}dn_{1}dn_{2},
\end{eqnarray}
where the total cross section of $\gamma \gamma\rightarrow e^{+}
e^{-}$ is\cite{36}
\begin{eqnarray}
\sigma_{\gamma\gamma}=\frac{4\pi\alpha^{2}}{M^{2}}\hat{\beta}_{L}\left[\frac{3-\hat{\beta}_{L}^{4}}{2\hat{\beta}_{L}}
\mathrm{ln}\frac{1+\hat{\beta}_{L}}{1-\hat{\beta}_{L}}-(2-\hat{\beta}_{L}^{2})\right],
\end{eqnarray}
and
\begin{eqnarray}
\hat{\beta}_{L}=\left[1-\frac{4m_{e}^{2}}{M^{2}}\right]^{1/2}.
\end{eqnarray}
In the semi-coherent case, $q_{1T}\gg q_{2T}$, the total transverse
momentum of dileptons is
$\vec{P}_{T}=\vec{q}_{1T}+\vec{q}_{2T}\approx \vec{q}_{1T}$, where
$q_{iT}$ is the transverse momentum of a photon. Therefore the
differential cross section can be written in the terms of dilepton
transverse momentum as the following
\begin{eqnarray}
\frac{d\sigma}{d^{2}p_{T}dMdy}&=&2\pi\left(\frac{2Z^{2}\alpha}{\pi}\right)^{2}\frac{1}{Z}\mathrm{ln}\frac{\gamma}{P_{T}R_{min}}
\frac{\sigma_{\gamma\gamma}}{p_{T}^{3}}\nonumber\\[1mm]
&&\times\int_{p_{2Tmin}}^{\gamma/R_{min}}\mathrm{ln}\frac{\gamma}{q_{2T}R_{min}}\frac{1}{q_{2T}}dq_{2T},
\end{eqnarray}
where the minimum transverse momentum of photon $q_{2}$ is
$q_{2Tmin}\sim 0.2$GeV due to the single track acceptance
condition. The factor $1/Z$ is the charge form factor. The form factor for low momentum coherent photon is $F^{2}(q_{T}\rightarrow 0)\sim1$. However, for the incoherent photon, the form factor is $F^{2}(q_{T}> 0)\approx \frac{1}{Z^{2}}Z$ \cite{33}. If both the photon transverse momenta ($q_{1T}$
and $q_{2T}$) are the same large (non-coherent) or small (coherent),
the total transverse momentum would have a very small value
($|\vec{P}_{T}|=|\vec{q}_{1T}-\vec{q}_{2T}|\rightarrow 0$), then a
dilepton could not obtain large transverse momentum in the $\gamma
\gamma\rightarrow e^{+} e^{-}$ interaction. The single track
condition $q_{2T}\geq 0.2$GeV allows us to discuss large transverse
momentum dilepton production in the two-photon reaction, the
contribution of non-coherent and coherent photon-photon processes is
weak compared with the semi-coherent processes. This is the reason
why the semi-coherent approach is essential in the two-photon
interactions.

\begin{figure}[t]
\includegraphics[width=9 cm]{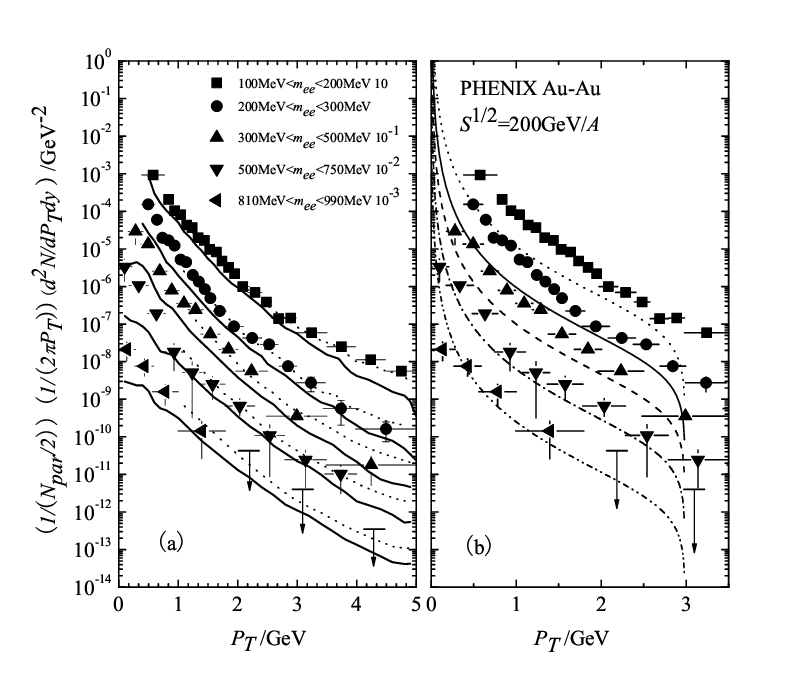}
\vspace{-4ex}\caption{ The cold component dilepton production via
the two-photon processes $\gamma\gamma\rightarrow e^{+}e^{-}$ for
different mass bins. (Panel a)The solid line: the sum of cocktail
and charm decay; the dot line: the sum of cocktail and charm decay
plus the contribution of direct dileptons [32]; (Panel b)the
contribution of $\gamma\gamma\rightarrow e^{+}e^{-}$, the dot line:
100MeV$<m_{ee}<$200MeV; the solid line: 200MeV$<m_{ee}<$300MeV; the
dash line: 300MeV$<m_{ee}<$500MeV; the dash dot line:
500MeV$<m_{ee}<$750MeV; the dash dot dot line:
810MeV$<m_{ee}<$990MeV. }\label{fig1}
\end{figure}

In the semi-coherent case ($q_{1}=(\omega_{1},q_{1T},q_{1Z})$ and
$q_{2}\simeq(\omega_{2},\vec{0},-q_{2Z})$), a non-coherent photon
with large transverse momentum is radiated from a nucleus and a
coherent photon with small transverse momentum is radiated from a
proton of another nucleus in the relativistic heavy ion collisions.
The large transverse momentum yield for different mass bins is given
as follows
\begin{eqnarray}
\frac{dN}{d^{2}p_{T}dy}=\int_{m_{1}}^{m_{2}}\frac{dN}{d^{2}p_{T}dMdy}dM,
\end{eqnarray}
where the yield relates to the differential cross section
$d\sigma/d^{2}p_{T}dMdy$ for nucleus-nucleus collisions with the
total cross section $\sigma_{tot}$ in the form
$dN/d^{2}p_{T}dMdy=(1/\sigma_{tot})(d\sigma/d^{2}p_{T}dMdy)$. The
authors in Ref\cite{33} use $\sigma_{tot}\sim 1.4\times
10^{4}\mathrm{GeV^{-2}}$ for the Au-Au 200GeV/A collisions at RHIC.

\section{Numerical results}
From Fig. 1 one can see that the contribution of two-photon
processes is only valid in the region of $P_{T}<$3GeV due to the
radiation limit condition $\omega<\gamma/R_{min}$. It also implies
that the spectra of $\gamma\gamma$ interaction depend on the nucleus
radiation energies. The energy of photons radiated from nuclei is
localized in the range of relatively small $q_{iT}$. Shuryak E. et.
al. have discussed the two-photon processes by using a charge
distribution form factor\cite{33}. A charge distribution is
necessary to the form factor, but the Woods-Saxon approximation they
used has depressed the value of the form factor, therefore the
spectra of dileptons under $\gamma\gamma\rightarrow e^{+}e^{-}$
interaction are also depressed quickly with the rising dilepton
transverse momentum. In this article we derive the yield of
dileptons for $\gamma\gamma$ processes with the restriction
$b\ll2R_{\perp}$, where the quark-gluon plasma may be created, and
it is found that the spectra for different mass bins are essential
in the soft region.

At low mass, $m_{ee}<$200MeV, the contribution of cocktail and charm
decays is in agreement with the PHENIX data. However, in the soft
region ($P_{T}<$1GeV) the data are still higher than the expectation
of cocktail and charm decays at $m_{ee}>$200MeV. The yield
enhancement is more evident with the increasing of dilepton mass. In
order to avoid the influence of $e^{+}e^{-}$ decays of narrow
resonance vector mesons, the mass regions around the $\omega$ meson
and $\phi$ meson are excluded by PHENIX Collaboration.

In Fig. 2 we plot the large transverse momentum dilepton yield under
$\gamma\gamma\rightarrow e^{+}e^{-}$ interactions for different mass
bins. As we discussed above the yield enhancement is not evident for
the mass bins as 100MeV$<m_{ee}<$200MeV and 200MeV$<m_{ee}<$300MeV,
so the correlation of cold $\gamma\gamma$ component is weak; In the
mass region of 300MeV$<m_{ee}<$500MeV, 500MeV$<m_{ee}<$750MeV and
810MeV$<m_{ee}<$990MeV the data are higher almost one order than the
expectation of cocktail and charm decays, and the modification of
$\gamma\gamma$ reaction is essential now. Therefore the two-photon
interaction plays an important role in the large transverse momentum
dilepton production in the soft $P_{T}$ region.

\begin{figure}[t]
\includegraphics[width=15cm]{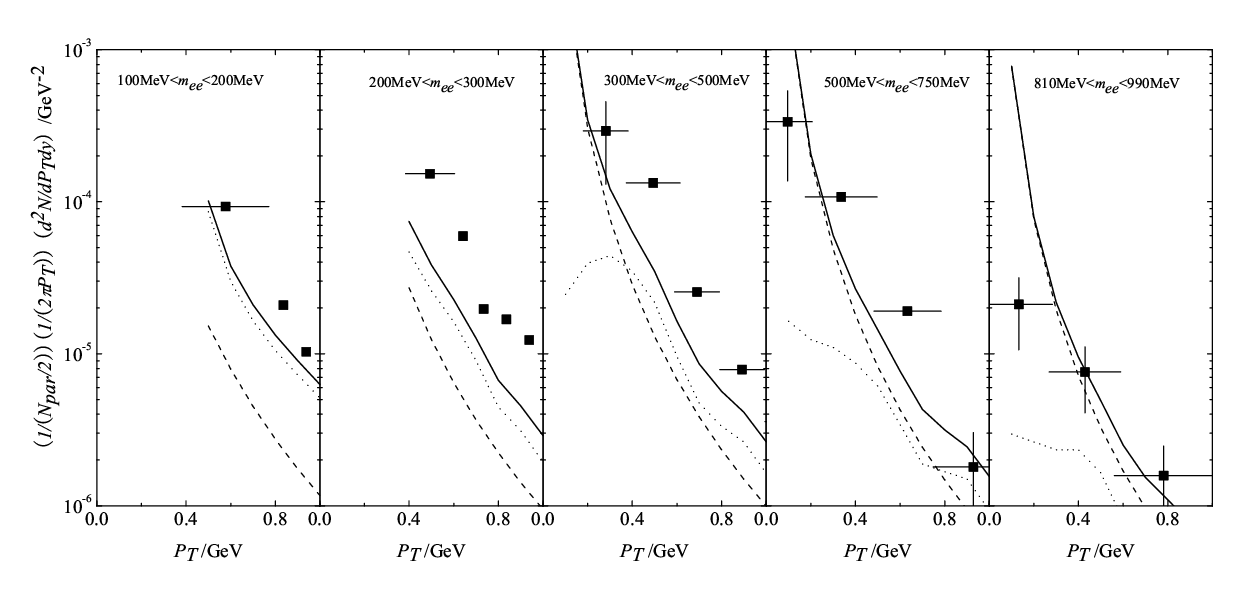}
\vspace{-4ex}\caption{ The dilepton spectra of two-photon processes
for different mass bins. The dilepton transverse momentum is in the
soft region $P_{T}<$1GeV. The dot line is the contribution of the
cocktail and charm decay, the dash line is the photon-photon
production, and the solid line is the total
contribution.}\label{fig2}
\end{figure}

\section{Conclusion}
We derived the two-photon processes from the passage of
$\gamma\gamma\rightarrow e^{+}e^{-}$. The large transverse momentum
dilepton can be created in the semi-coherent two-photon interaction.
The $\gamma\gamma$ processes may be an essential complementarity to
the previous standard cold dilepton production. In the MacDonald
approximation the energy of the radiated photon is limited in the
region of $\omega <\gamma/R_{min}$, therefore the cold dilepton
component of two-photon processes is still valid in the soft $P_{T}$
region where the pQCD is out of use.

The numerical results are plotted by comparing with the PHENIX
experimental data. The modification of two-photon processes is more
evident with the rising dilepton mass, especially in the mass bins
of 300MeV$<m_{ee}<$500MeV, 500MeV$<m_{ee}<$750MeV and
810MeV$<m_{ee}<$990MeV.

This work is supported by the National Natural Science Foundation of
China (10665003 and 11065010).


\end{document}